\documentclass[3p,times,twocolumn]{elsarticle}

\usepackage{ecrc, color}

\newcommand{\br}{{\rm BR}}
\newcommand{\fb}{\ensuremath{{\rm fb}^{-1}}}
\newcommand{\gmt}{\ensuremath{(g-2)_\mu}}

\newcommand{\bsmm}{\ensuremath{\br(B_s \to \mu^+\mu^-)}}
\newcommand{\msq}{m_{\tilde{ q}}}

\newcommand{\mgl}{\ensuremath{m_{\tilde g}}}
\newcommand{\mneu}[1]{\ensuremath{m_{\tilde \chi^0_{#1}}}}
\newcommand{\mst}[1]{\ensuremath{m_{\tilde t_{#1}}}}

\newcommand{\Mh}{\ensuremath{M_h}}

\newcommand{\MA}{\ensuremath{M_A}}
\newcommand{\tev}{\ensuremath{\,\, \mathrm{TeV}}}
\newcommand{\gev}{\ensuremath{\,\, \mathrm{GeV}}}

\newcommand{\ssi}{\ensuremath{\sigma^{\rm SI}_p}}
\newcommand{\tb}{\ensuremath{\tan\beta}}
\newcommand{\msqfs}{\ensuremath{m_{\widetilde{q}_{12}}}}
\newcommand{\msqt}{\ensuremath{m_{\widetilde{q}_3}}}
\newcommand{\msl}{\ensuremath{m_{\widetilde{l}}}}
\newcommand{\At}{\ensuremath{A_t}}
\definecolor{orange}{rgb}{1,0.5,0}

\newcommand{\ETslash}{/ \hspace{-.7em} E_T}

\volume{00}

\firstpage{1}

\journalname{Nuclear Physics B Proceedings Supplement}

\runauth{}

\jid{nuphbp}

\jnltitlelogo{Nuclear Physics B Proceedings Supplement}

\usepackage{amssymb}
\biboptions{sort&compress}

\usepackage[figuresright]{rotating}

\begin{document}

\begin{frontmatter}



\dochead{}

\title{SUSY fits with full LHC Run I data}


\author[ic]{Kees Jan de Vries (on behalf of the MasterCode Collaboration)}

\address[ic]{High Energy Physics Group, Blackett Laboratory, Imperial College, 
             Prince Consort Road, London SW7 2AZ, UK}

\begin{abstract}
We present the latest results from the MasterCode collaboration on 
supersymmetric models, in particular on the CMSSM, 
the NUHM1, the NUHM2 and the pMSSM. We combine the data from LHC Run I with 
astrophysical observables, flavor and electroweak precision observables. 
We determine the best fit regions of these models and analyze the discovery 
potential of squarks and gluinos at LHC Run II and direct detection experiments.
\end{abstract}

\begin{keyword}
Supersymmetry \sep CMSSM \sep NUHM1 \sep NUHM2 \sep pMSSM
\end{keyword}

\end{frontmatter}

% \linenumbers

\section{Introduction}

Despite the absence of any convincing signal of supersummetry (SUSY) after Run 1 
at the large hadron collider (LHC), SUSY remains well motivated. 
First of all, the lightest neutralino is a natural DM candidate.
Secondly, SUSY provides a solution to the hierarchy problem.
Finally, SUSY allows for unification of the gauge coupling at the so-called grand 
unified theory (GUT) scale of $\mathcal{O}(10^{16}\gev)$. 

In these proceedings we present a selection of the results from global
frequentist fits of constrained models of SUSY - the CMSSM, NUHM1, 
NUHM2 and pMSSM10 (defined below) -
to experimental constraints from Run I LHC data,
astrophysical observables, flavor and electroweak observables. 
The fits allow us to identify the relevant parameters, assess and compare the 
validity of the models and study the predictions and consequences for future
searches and experiments. 
In particular, we focus on the differences between GUT-scale and
phenomenological models highlighting the \gmt\ constraint. 
We discuss the discovery potential for gluinos and
squarks at LHC Run II as well as prospects for direct detection of dark matter.

Note that the results presented in these proceedings date from the ICHEP2014
conference.  
We have published elsewhere some of the results shown in these proceedings, namely on the CMSSM, 
NUHM1 and NUHM2~\cite{Buchmueller:2013rsa,Buchmueller:2014yva}. 
We would also like to mention that there are several other groups that perform
global fits of SUSY using Bayesian as well as frequentist methods. 
Some recent fits of CMSSM, NUHM1 and NUHM2 may be found
in~\cite{Bechtle:2012zk, Scott:2012mq, Strege:2012bt, Roszkowski:2014wqa},
whereas results on the pMSSM may be found in ~\cite{Boehm:2013qva,Strege:2014ija}.
We will soon publish updated results on pMSSM10. 

\section{Analysis procedure}
\label{sec:procedure}

\subsection{Models}

We consider four constrained versions of the general $R$-parity-conserving
Minimal Supersymmetric extension of the Standard Model (MSSM). 
Three of these models are derived from GUT
model-building considerations, where masses and couplings are assumed to unify at
the GUT scale: 
In the constrained MSSM (CMSSM) 
all scalars (two Higgs doublets and the sfermions) have a universal soft SUSY-breaking mass $m_0$, 
the gauginos a universal mass $m_{1/2}$, 
and the trilinear couplings are all equal to  $A_0$. 
In the NUHM1 the masses of the Higgs doublets are assumed to be independent but 
equal, while in the NUHM2 they are allowed to vary independently. 
In general $m_0^2$ can take negative values, and so we denote in this paper 
$m_0 \equiv {\rm Sign}(m_0^2) \sqrt{|m_0^2|} < 0$.
The remaining parameters of these models are the superpotential coupling $\mu$ between
the Higgs doublets and the ratio of the vacuum expectation
values of the two Higgs doublets, $\tan\beta\equiv v_1/v_2$. 
We also consider a 10-dimensional subset of the 
so-called phenomenological MSSM (pMSSM)\cite{Djouadi:1998di}, which makes no 
assumptions about the masses at the GUT-scale. 
Instead in the pMSSM the soft SUSY-breaking parameters are defined at the 
SUSY-breaking scale $M_{SUSY}\sim\sqrt{\mst{1}\cdot\mst{2}}$. 
It also is assumed there are no flavor-changing neutral currents, 
no additional sources of CP violation as well as unification of the first and second 
generation sfermion masses.
Our 10-dimensional subset of the pMSSM (pMSSM10) is defined as follows.
We set all first and second generation squark masses to a common value $\msqfs$, 
all third-generation squark mass parameters to a common value $\msqt$,  
the slepton masses to $\msl$, and the trilinear couplings $ \At = A_b =
A_{\tau} = A$. 
The remaining parameters are the gaugino masses, $M_1,~M_2,~M_3$, the Higgs mixing
parameter $\mu$, the CP-odd Higgs mass scale $\MA$,  and $\tb$. 

\subsection{Fitting procedure}
We construct a $\chi^2$ function in the same way as
in~\cite{Buchmueller:2014yva,Feroz:2011bj}, taking into account 
constraints from  B-physics, electroweak precision observables, 
cosmology and direct SUSY searches at the LHC. 
The only difference of the $\chi^2$ function between the models is the way we
implement the direct SUSY searches. 
For the CMSSM, NUHM1 and NUHM2 we use the latest results from the jets + 
$\ETslash$ analysis of ATLAS~\cite{Aad:2014wea} and apply the constraint as we 
previously described in~\cite{Buchmueller:2013rsa}.
For the pMSSM10 we use a 4-dimensional lookup table in (\mneu, \mgl, \msqfs, \msqt).
The underlying principle of this method is that, to good approximation, the
limits from direct SUSY searches are independent of the configuration of the
sleptons and the other gauginos as was argued in~\cite{Buchmueller:2013exa}.
We use the same framework as was used for that work, using 7\tev\ searches of 
CMS~\cite{Chatrchyan:2012wa,Chatrchyan:2012ola,Chatrchyan:2012te,Chatrchyan:2012sa}.
We have extensively validated this approach but details are beyond the scope of 
these proceedings, and will be presented in future work. 

Finally, we use the MasterCode framework 
to calculate the observables that go into the $\chi^2$ calculation. 
The MasterCode framework interfaces various public and private codes using the
SLHA format~\cite{Allanach:2008qq}. 
In particular we use \texttt{SOFTSUSY}~\cite{Allanach:2001kg} to calculate the spectra,
FeynHiggs~\cite{Hahn:2013ria} to calculate Higgs observables and \gmt,
\texttt{MicrOMEGAs}~\cite{Belanger:2013oya} to calculate the relic DM density,
SuFla~\cite{Isidori:2006pk,Isidori:2007jw} for
B-physics observables, FeynWZ~\cite{Heinemeyer:2006px,Heinemeyer:2007bw} for EWPOs, and SSARD for the
spin-independent cross section.
We use the Multinest package~\cite{Feroz:2008xx} for sampling. 

\section{Results}

\subsection{$(m_0,m_{1/2})$ plane of the CMSSM, NUHM1 and NUHM2}

%%%%%%%%%%%%%%%%%%% Figure %%%%%%%%%%%%%%%%%%
\begin{figure}[htb!]
\includegraphics[width=0.5\textwidth]{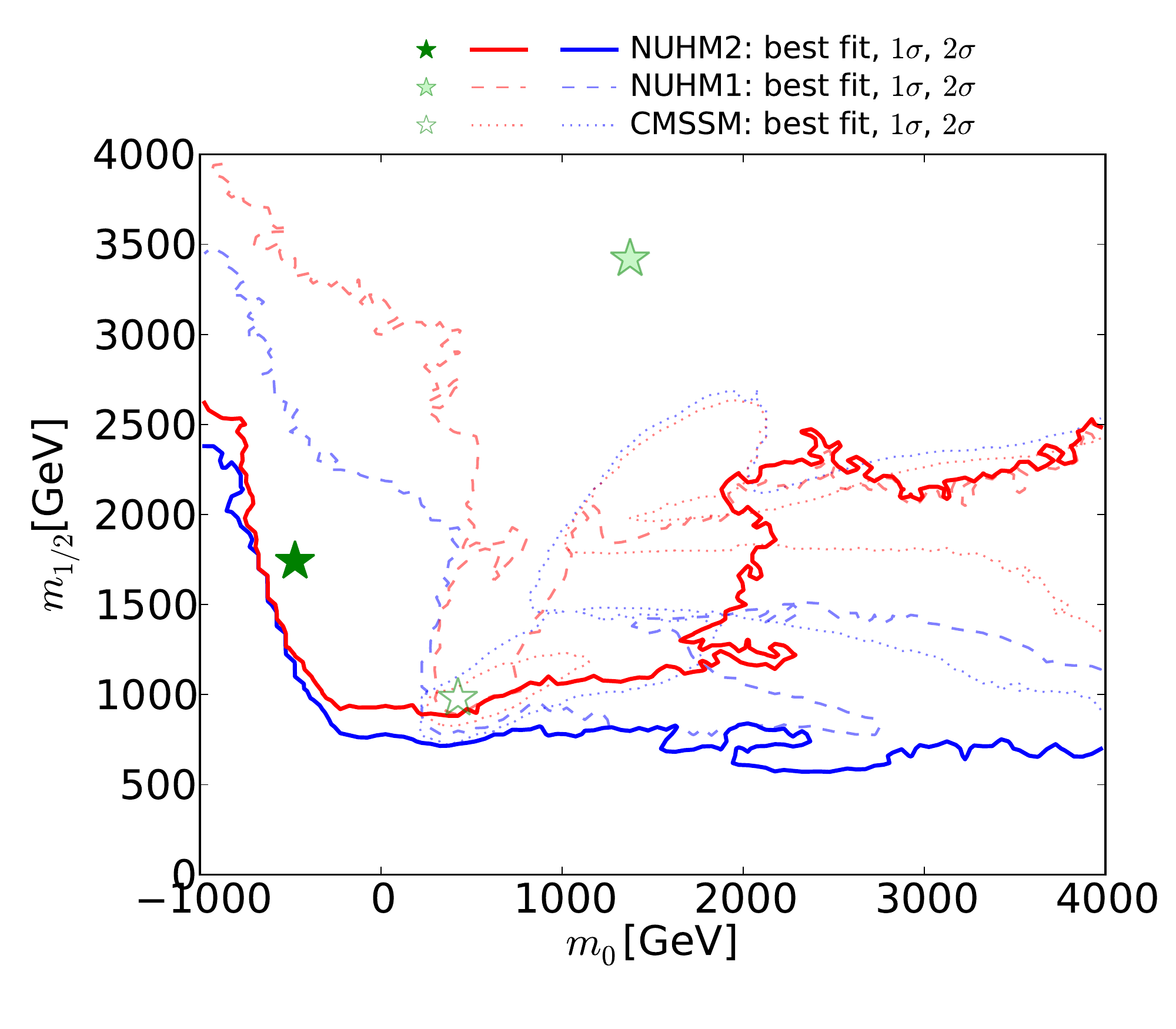}
\caption{The $(m_0,m_{1/2})$ parameter plane of the CMSSM, NUHM1 and NUHM2. Red
 (blue) dotted, dashed and solid contours correspond to their respective 68\% 
(95\%) CL, whereas empty, shaded green and filled stars correspond to their 
respective best fit points.}
\label{fig:m0m12}
\end{figure}
%%%%%%%%%%%%%%%%%%% Figure %%%%%%%%%%%%%%%%%%

Let us first turn to the $(m_0,m_{1/2})$ parameter plane of the CMSSM, NUHM1 
and NUHM2 in Fig.~\ref{fig:m0m12}. In this figure the red (blue) dotted, dashed,
and solid lines correspond to the 68\% (95\%)~CL contours of the CMSSM, NUHM1 
and NUHM2 respectively. The empty, light shaded green and green filled stars 
denote the respective best fit points.

In the CMSSM there appears a bimodal structure, with one local minimum around 
$m_0 \sim 400$ and $m_{1/2} \sim 1000$~GeV and another that stretches to high 
values of $m_0$ at high $m_{1/2}$. 
These modes correspond to two different mechanisms to fulfil the relic DM
density constraint, namely stau coannihilation and 
the heavy Higgs ($H/A$) funnel respectively.
We see that in the displayed region of the CMSSM, the $m_{1/2}$ parameter has an upper bound of
approximately 2.5~TeV,
it should be stressed that this lower bound increases for larger values of $m_0$.
The lower bound on $m_{1/2}$ at $m_0\sim2\tev$ is mainly due to the \bsmm\ constraint. 
From our previous fits \cite{Buchmueller:2012hv,Buchmueller:2013rsa} 
we know that the \gmt\ constraint prefers low values of 
$m_0$ and $m_{1/2}$, which are in tension with the absence of any signal from 
searches for SUSY particles at the LHC.

In the NUHM1 we note that there is no longer an upper bound on $m_{1/2}$.
The region that has become available at $m_{1/2}\gtrsim2.5\tev$ is characterized
by chargino coannihilation. 
We also see that in the NUHM1 negative values of $m_0$ are accessible at 
$m_{1/2} \gtrsim 2.2\tev$ at 95\%~CL.
In this region both stau and chargino coannihilation are responsible for
fulfilling the relic DM density constraint.
Stau coannihilation and the $H/A$ funnel appear at similar places as in the
CMSSM.
In the NUHM2, the extra degree of freedom allows the stau coannihilation region
to expand to negative values of $m_0$ at low values of $m_{1/2}$.

We would like to emphasize that in general the 68\% and 95\%~CL contours extend
beyond the boundary of the sampled parameter ranges. 
This highlights the fact that the minimum structure is very shallow and there is
no particularly favored region in the parameter space.

\subsection{The anomalous magnetic dipole moment of the muon \gmt\ }

There is a discrepancy of $\sim3.5\sigma$ between the 
measurement~\cite{Bennett:2006fi} and 
the theoretical SM calculation~\cite{Benayoun:2012wc} (and references therein) 
of the \gmt. 
This discrepancy can be interpreted as arising from SUSY 
contributions, see e.g.~\cite{Stockinger:2006zn}.
Sizable SUSY contributions can arise when neutralinos, 
charginos, smuons and muon sneutrinos have masses of $\mathcal{O} (100\gev)$.
However, one has to keep in mind that in GUT-models the chargino and neutralino 
masses are directly proportional 
to the gluino mass, and are hence directly constrained by gluino searches at the 
LHC. 
Similarly, the smuon and muon sneutrino masses are constrained 
by searches for colored sparticles at the LHC. 
These searches have in general a greater sensitivity than searches for 
electroweakly interacting sparticles.

%%%%%%%%%%%%%%%%%%% Figure %%%%%%%%%%%%%%%%%%
\begin{figure}[htb!]
\includegraphics[width=0.5\textwidth]{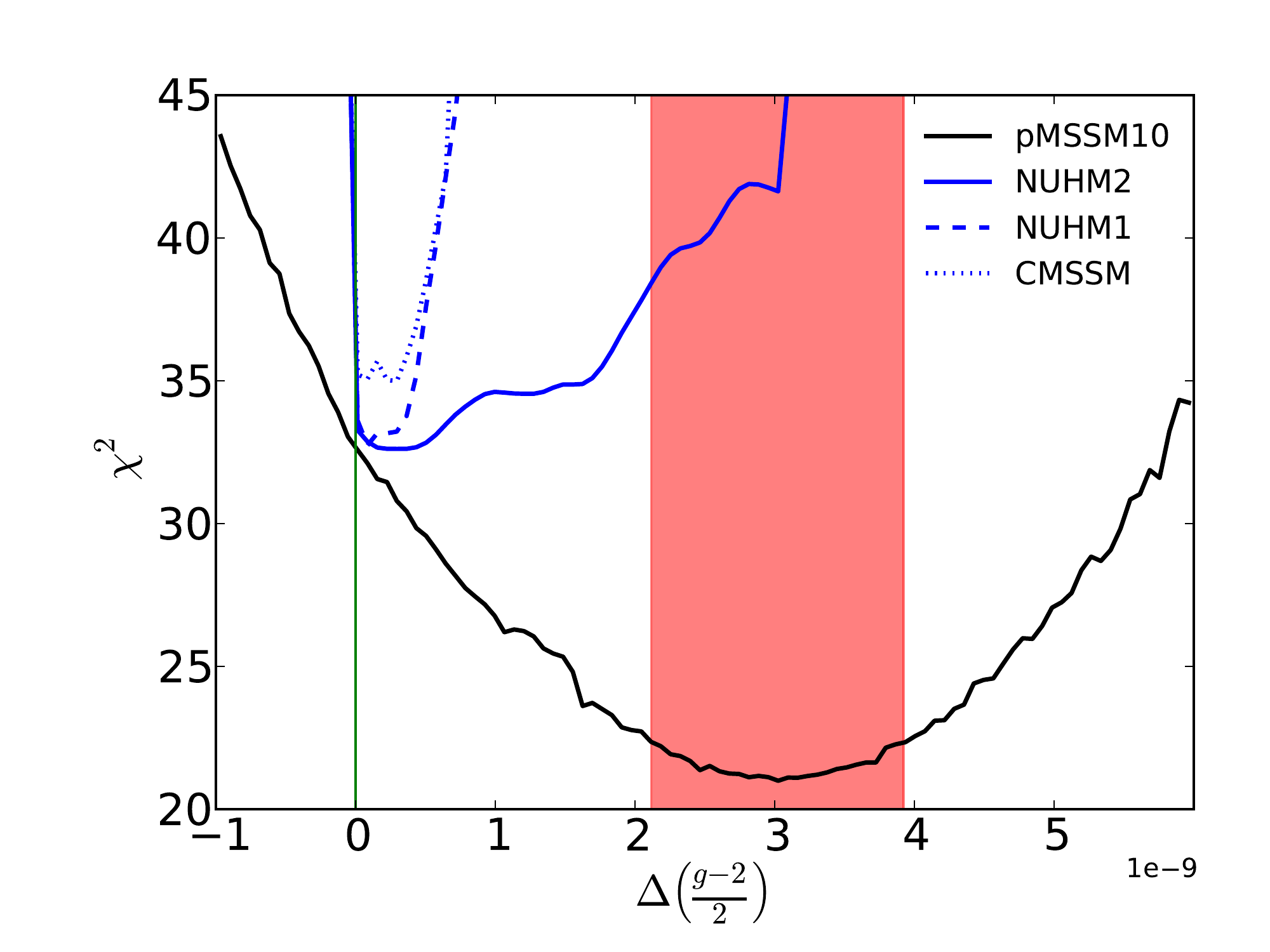}
\caption{Total $\chi^2$ distribution for SUSY contribution to the anomalous 
magnetic dipole moment of the muon $\Delta \left(\frac{g-2}{2} \right)$. The green
line corresponds to the SM value, whereas the red band corresponds to the 
measured value and its uncertainty.}
\label{fig:g-2}
\end{figure}
%%%%%%%%%%%%%%%%%%% Figure %%%%%%%%%%%%%%%%%%
In global fits of the CMSSM and NUHM1\cite{Buchmueller:2013rsa} we 
found that indeed \gmt\ cannot be
reconciled with the non-observation of sparticles at the LHC. 
The question is whether the additional freedom in NUHM2 and pMSSM10 allows for 
\gmt\ to be fulfilled. 
Fig.~\ref{fig:g-2} displays the total $\chi^2$ as a function of half the 
difference between the predicted value of \gmt\ and the SM value
$\Delta\left(\frac{g-2}{2} \right)$ for the CMSSM, NUHM1, NUHM2 and pMSSM10 in
dotted blue, dashed blue, solid blue and solid black respectively. 
To guide the eye a green vertical line is displayed to indicate where \gmt\
equals the SM value. 
A shaded red band indicates the current experimental value.

We see that in the CMSSM and NUHM1 \gmt\ cannot be fulfilled. 
Other constraints, in particular ATLAS jets + $\ETslash$ search, 
force \gmt\ to take values close to the SM prediction. 
In the NUHM2, however,  we see that we can indeed get \gmt\ at the measured value, 
although at the expense of other constraints,  notably \Mh~ and the ATLAS constraint.
In the pMSSM10 the tension is completely lifted. 

%%%%%%%%%%%%%%%%%%%%%% T A B L E %%%%%%%%%%%%%%%%%%%%%%%%%%%%%%%%%%%%%%%%%
\begin{table*}[htb!]
\begin{center}
\begin{tabular}{lcc} \hline
Model   & $\chi^2$/d.o.f. & $p$-value  \\ 
\hline   
CMSSM   & 35.0/23         & 5.2  \%  \\
NUHM1   & 32.7/22         & 6.6  \%  \\
NUHM2   & 32.5/21         & 5.2  \%  \\
pMSSM10 & 21.3/17         & 21   \%  \\
\hline
\end{tabular}
\caption{Minimum values of the total $\chi^2$ over the number of degrees of 
         freedom and the corresponding $p$-value for each model model.} 
\label{tab:bestfits}
\end{center}
\end{table*}
%%%%%%%%%%%%%%%%%%%%%% T A B L E %%%%%%%%%%%%%%%%%%%%%%%%%%%%%%%%%%%%%%%%

Table~\ref{tab:bestfits} shows the total $\chi^2$ over the number of degrees of 
freedom and the corresponding $p$-values for the CMSSM, NUHM1, NUHM2 and pMSSM10. 
We note that the total $\chi^2$ drops by more than 10 units when
going from the NUHM2 to the pMSSM10. 
Correspondingly, the $p$-value increases from $\sim5.2\%$ to $\sim 21\%$
indicating a significant improvement of the fit. 
From Fig.~\ref{fig:g-2} it is obvious that the greatest improvement is gained
from reconciling the SUSY interpretation of \gmt\ with the other constraints. 
Hence, the pMSSM10 does indeed allow  the tension
between \gmt\ and other constraints to be lifted.

\subsection{LHC Run 2 discovery potential of squarks and gluinos}

%%%%%%%%%%%%%%%%%%% Figure %%%%%%%%%%%%%%%%%%
\begin{figure}
\includegraphics[width=0.5\textwidth]{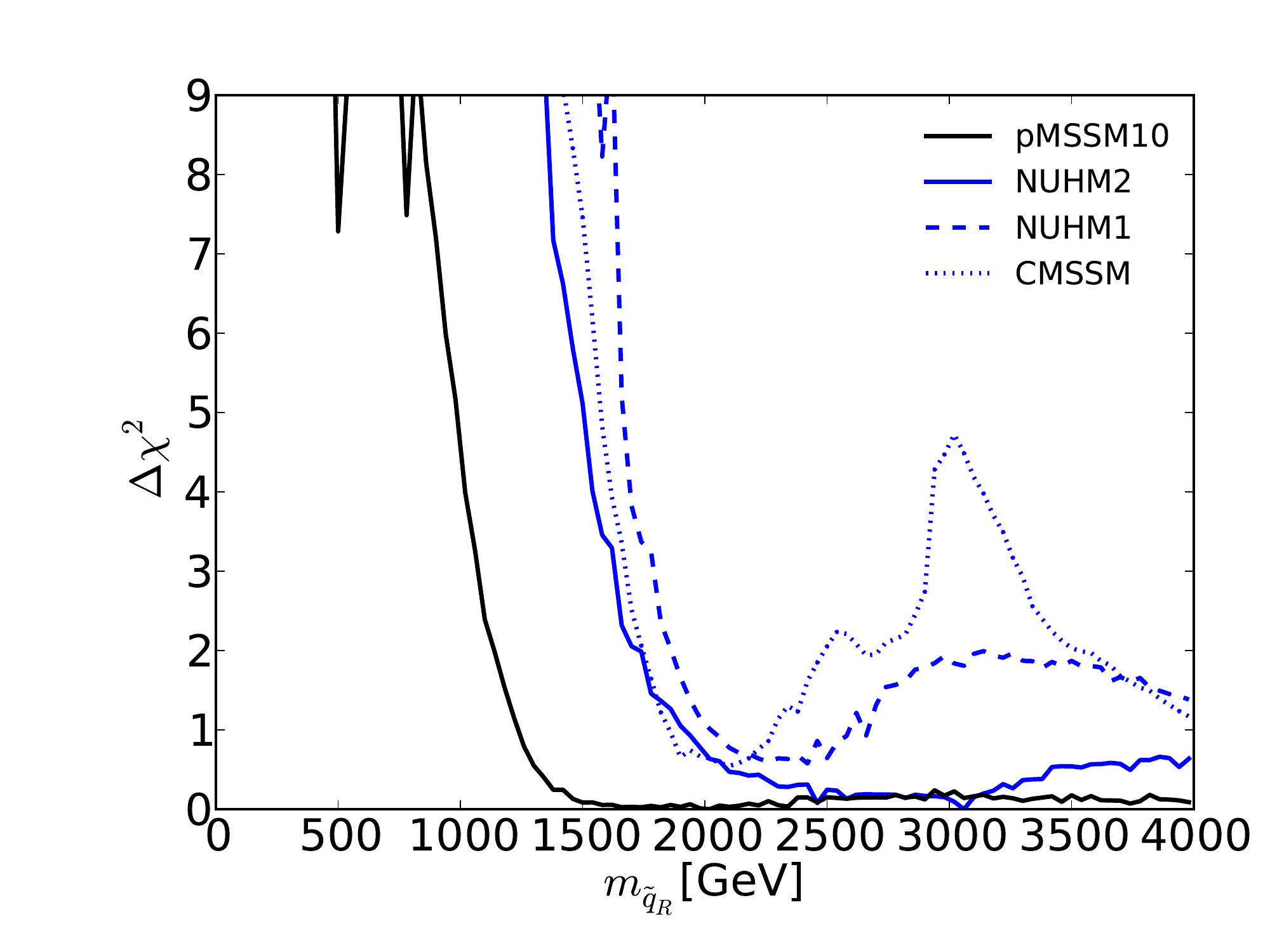}
\caption{$\Delta\chi^2$ distribution for the first and second generation 
(right-handed) squark masses. The blue dotted, blue dashed, blue solid and 
black solid lines correspond to CMSSM, NUHM1, NUHM2 and pMSSM10 respectively. 
Note that the pMSSM10 results are only for LHC7 data.}
\label{fig:msqr}
\end{figure}
%%%%%%%%%%%%%%%%%%% Figure %%%%%%%%%%%%%%%%%%

%%%%%%%%%%%%%%%%%%% Figure %%%%%%%%%%%%%%%%%%
\begin{figure}
\includegraphics[width=0.5\textwidth]{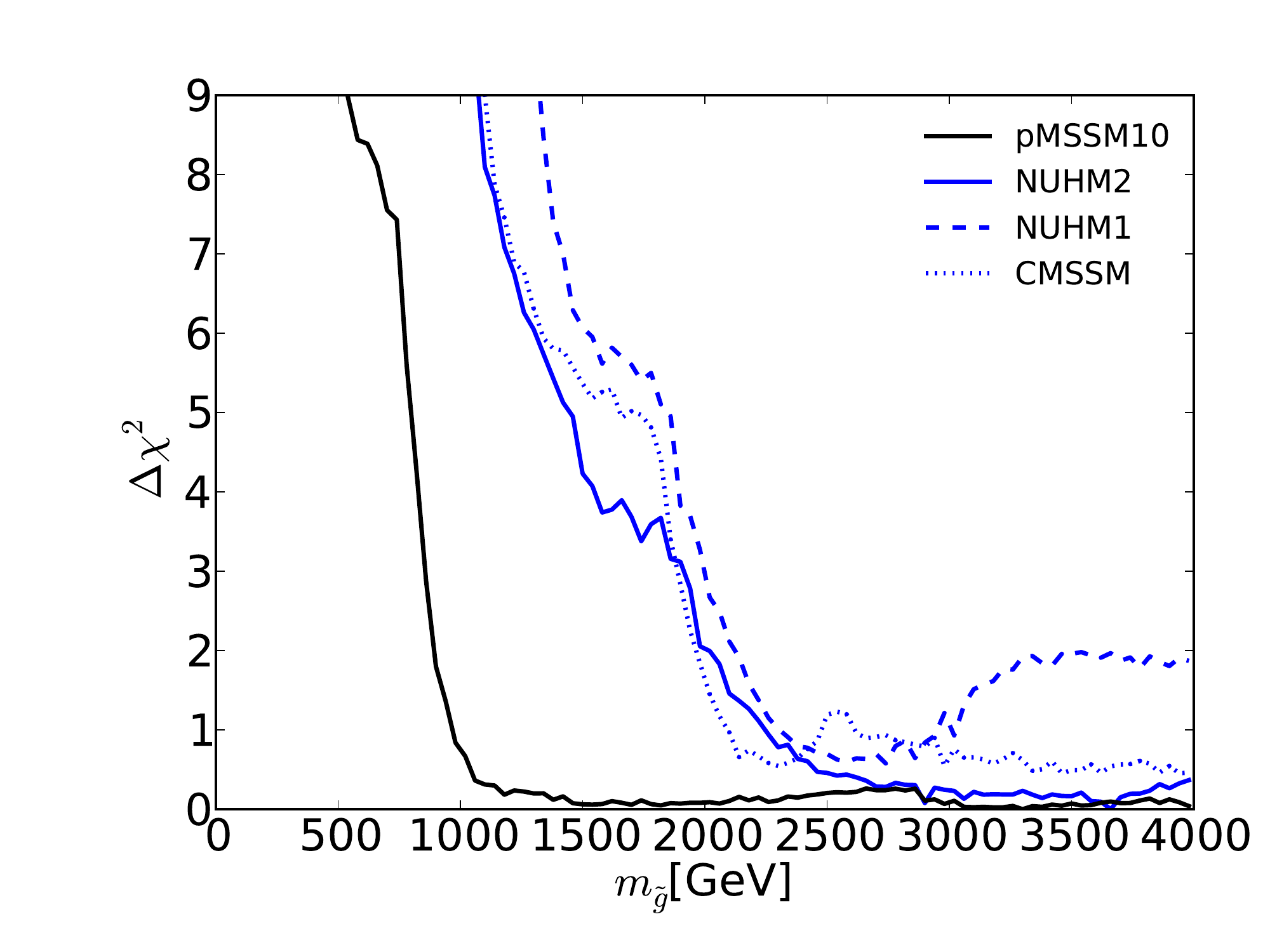}
\caption{$\Delta\chi^2$ distribution for the gluino mass. The blue dotted, blue 
dashed, blue solid and 
black solid lines correspond to CMSSM, NUHM1, NUHM2 and pMSSM10 respectively. 
Note that the pMSSM10 results are only for LHC7 data.}
\label{fig:mg}
\end{figure}
%%%%%%%%%%%%%%%%%%% Figure %%%%%%%%%%%%%%%%%%

We now turn to the one-dimensional profile likelihoods for the right-handed first-
and second-generation squark and gluino masses in Fig.~\ref{fig:msqr} and 
Fig.~\ref{fig:mg} respectively. 
In these figures the CMSSM, NUHM1, NUHM2 and pMSSM10 results are indicated by dotted
blue, dashed blue, solid blue and solid black respectively.
We first observe that lower bounds on the gluino and the squark masses in the
CMSSM, NUHM1 and NUHM2 are almost identical, namely $\sim1.7\tev$  
for the right handed squarks and $\sim1.8\tev$ for the gluinos 
at $\Delta\chi^2\sim4$. 
In the CMSSM the bimodal structure is again visible for the squark
mass. 
To a lesser extent this is true for the NUHM1 and NUHM2. 
The lower bounds in the pMSSM10 are significantly lower, which is expected
because only the 7\tev\ limits have been applied. 

The predictions for the squarks and gluinos may be compared (with some caveats) 
to a study by the ATLAS collaboration~\cite{atlas_simulation}, Fig.~5(a). 
According to this figure, the exclusion potential after 300~\fb at $14\tev$ is 
$\sim2.7\tev$ ($\sim2.3\tev$) for $\msq$ (\mgl), irrespective of \mgl ($\msq$), when
assuming $\mneu{1}=0\gev$. 
According to these numbers, the lower-mass mode of the CMSSM should be
accessible in LHC Run II. 
However, masses higher than $\sim3\tev$ might be beyond reach even at high
luminosity LHC. 

%%%%%%%%%%%%%%%%%%% Figure %%%%%%%%%%%%%%%%%%
\begin{figure}
\includegraphics[width=0.5\textwidth]{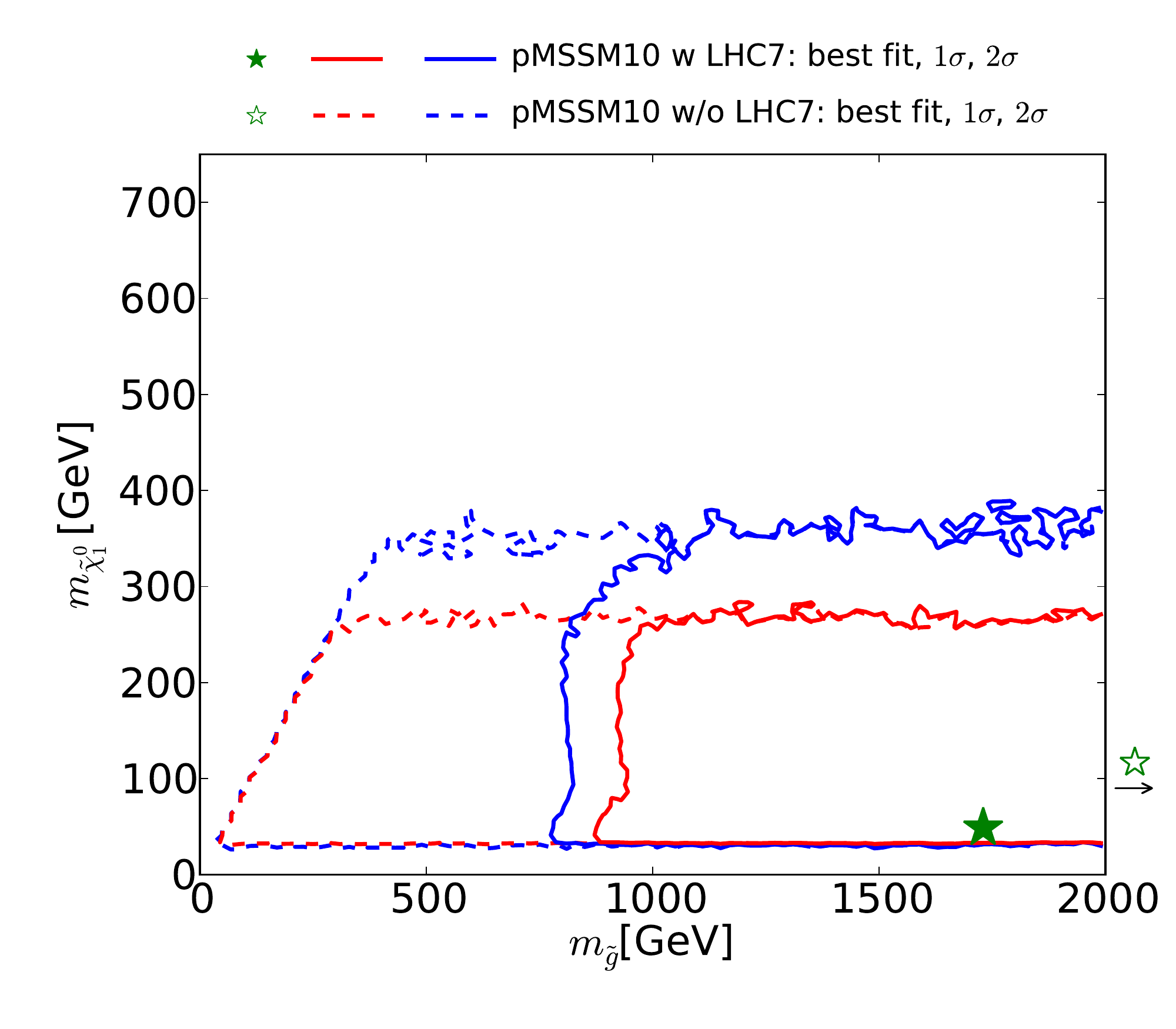}
\caption{The $(\mgl, \mneu{1})$ plane in the pMSSM10, with and without the 
colored searches at 7\tev\ (indicated as LHC7) applied. 
The red (blue) solid and dashed contours correspond to 68\% (95\%)~CL
respectively, whereas the filled and empty star correspond to the respective 
best fit points. 
Note that the best fit point without LHC7 applied lies outside the plotted range. 
An arrow indicates its position.}
\label{fig:msqrmneu}
\end{figure}
%%%%%%%%%%%%%%%%%%% Figure %%%%%%%%%%%%%%%%%%

In  Fig.~\ref{fig:msqrmneu} we display the $(\mgl,\mneu{1})$ plane for
the pMSSM10. 
In this figure the solid and dashed red (blue) contours correspond to 68\% 
(95\%)~CL of the pMSSM10 with and without the LHC searches for sparticles
applied respectively. 
The filled and the empty star correspond to their respective best-fit points.
A very important feature is that in the pMSSM10 the gluino and neutralino mass
are completely independent, as opposed to GUT-scale models which have a fixed
relationship between these masses.  
This freedom corresponds nicely to possibilities studied by the CMS and ATLAS experiments
in simplified models~\cite{Alves:2011wf}. 
We also see that values $\mneu{1}\lesssim300$ are preferred. 
This strong preference can be understood from the fulfillment of the \gmt\
constraint.
The gluino mass is generally unconstrained once it is above the CMS-imposed lower bound. 

\subsection{Direct Dark Matter detection}

\begin{figure}
\includegraphics[width=0.5\textwidth]{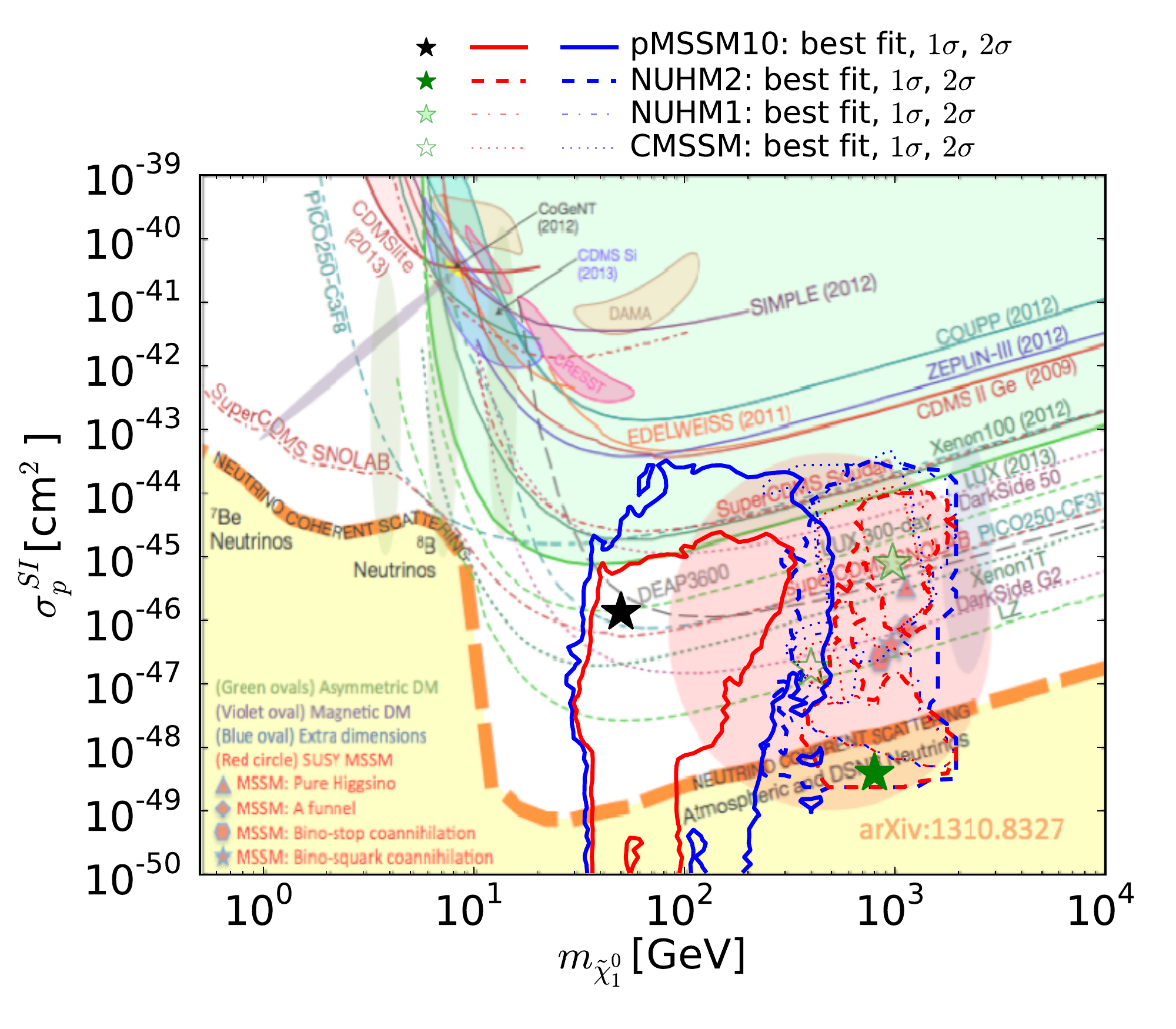}
\caption{Our results for $(\mneu{1}, \ssi)$ plane of the CMSSM, NUHM1, NUHM2 and pMSSM10
overlaid on the summary plot from Snowmass CF1 Summary~\cite{Cushman:2013zza}. 
The description of the summary plot can be found in the text. 
The overlaid red (blue) dotted, dashed-dotted, dashed and solid correspond to 
the 68\% (95\%)~CL for the CMSSM, NUHM1, NUHM2 and pMSSM10 respectively, whereas the empty, shaded green, green filled 
and black stars correspond to their best fit points.
}
\label{fig:mlspssi}
\end{figure}

Finally we turn to predictions for the spin-independent cross-section as a
function of the lightest neutralino mass in Fig.~\ref{fig:mlspssi}. 
Here we overlaid our results on the summary plot of the Snowmass CF1 Summary
~\cite{Cushman:2013zza}. 
The summary plot shows current upper bounds on the spin-independent cross
section as a function of the lightest neutralino mass
from various experiments using solid
lines. 
Dashed lines indicate the projected sensitivity of future 
searches. 
The yellow region indicates where backgrounds from solar and atmospheric
neutrinos dominate over the DM signal. 
We will refer to this boundary as the neutrino floor. 
Our results are shown in solid, dashed-dotted, dashed and dotted red (blue) 
contours corresponding to 68\% and (95\%)~CL for the pMSSM10, NUHM2, NUHM1 and 
CMSSM respectively.
The black, green, light shaded green and empty stars correspond to their best fit
points respectively.

We see that future experiments would probe a significant part of the parameter
space of the CMSSM, and all of the preferred regions are above the 
neutrino floor. 
For the NUHM1 some of the allowed region is below the neutrino
floor, whereas the NUHM2 even has its best-fit point in this region.
In the pMSSM10, the preferred masses of the lightest
neutralino are significantly lower than those in the CMSSM, NUHM1 or NUHM2.
Another striking feature of the pMSSM10 is that the cross sections can go down
to the extremely low value of $\ssi \ll 10^{-50}$, which would make it very hard
if not impossible for direct detection experiments to measure dark matter
scattering.

%%%%%%%%%%%%%%%%%%% Figure %%%%%%%%%%%%%%%%%%
%%%%%%%%%%%%%%%%%%% Figure %%%%%%%%%%%%%%%%%%

\section{Conclusion}

In supergravity-inspired models like the CMSSM, NUHM1 and NUHM2, there has been an ever 
growing tension between a supersymmetric interpretation of \gmt\ and the searches
for SUSY particles and other new phenomena at the LHC during Run~1. 
We showed that even the extra freedom in the NUHM2 cannot resolve this tension.
However, in the pMSSM10, where no GUT-scale unifying assumptions are made, 
it is indeed possible to reconcile \gmt\ with the measurements and
non-measurements at the LHC. 
This significantly improves the fit. 
In particular, the $p$-value increases from $\sim5\%$ to $\sim20\%$ when going
from the SUGRA models to pMSSM10.

We discussed the discovery potential of the first two generations (right-handed)
squarks and gluinos at LHC Run II.
The CMSSM, NUHM1 and NUHM2 have very similar lower limits of $\sim1.7\tev$  
for the squarks and $\sim1.8\tev$ for the gluinos. 
The lower bounds for the pMSSM10 are lower, but one has to keep in mind that we
only applied 7\tev\ limits. 
Comparing to predicted sensitivities from ATLAS~\cite{atlas_simulation}, 
it seems that the mode
at low mass in the CMSSM will be within reach of LHC Run II. 
The other models also have good parameter space within reach. 
However, the gluino and squark masses are genenerally unconstrained when above
the ATLAS/CMS-imposed lower bounds. 
Nevertheless, the pMSSM10 already provides complementary information compared to
GUT models, as the relations between the gluino and neutralinos are completely
lifted.

Finally we saw that  future dark matter direct detection experiments 
will have access to a significant part of the parameter space of the
CMSSM, NUHM1 and NUHM2. 
However some of the preferred regions for the latter two lie in the region where
the background from atmospheric neutrinos dominates. 
In the pMSSM10, very low spin-independent cross sections are possible. 
Scattering cross-sections below the neutrino floor might render it very hard if 
not impossible for direct detection experiments to measure any signal.



%% References

%\nocite{*}
\bibliographystyle{elsarticle-num}
\bibliography{martin}

%% The provided style file elsarticle-num.bst formats references in the 
%% required Procedia style


%\begin{thebibliography}{00}


%\end{thebibliography}

\end{document}